\documentclass[11pt,twoside]{article}

\usepackage{asp2006}

\begin{document}
\title{Secular Evolution in Disk Galaxies: Pseudobulge Growth\\ and the Formation of Spheroidal Galaxies} 
\author{John Kormendy and David B.~Fisher}   
\affil{Department of Astronomy, University of Texas at Austin, 1 University {\phantom{000}}Station C1400, 
       Austin, TX 78712-0259, USA;\\
       MPI f\"ur Extraterr.~Physik, Postfach 1312, 85741 Garching, Germany;\\
       Universit\"ats-Sternwarte, Scheinerstrasse 1, 81679 Munich, Germany}    

\begin{abstract} 
\pretolerance=15000  \tolerance=15000
\def\gapprox{$_>\atop{^\sim}$} 
\def\lapprox{$_<\atop{^\sim}$}
Updating Kormendy \& Kennicutt (2004, ARA\&A, 42, 603), we review internal secular 
evolution of galaxy disks.  One consequence is the growth of pseudobulges that often 
are mistaken for true (merger-built) bulges.  Many pseudobulges are recognizable 
as cold, rapidly rotating, disky structures.  Bulges have S\'ersic function brightness 
profiles with index 
$n$ \gapprox \thinspace2 while most pseudobulges have $n$ \lapprox \thinspace2.  
Recognition of pseudobulges makes the biggest problem with cold dark matter galaxy 
formation more acute: How can hierarchical clustering make so many pure disk galaxies
with no evidence for merger-built bulges?  E.{\thinspace}g., the giant Scd galaxies
M{\thinspace}101 and NGC 6946 have rotation velocities of $V_{\rm circ} \sim 200$ km~s$^{-1}$
but nuclear star clusters with velocity dispersions of 25 to 40 km s$^{-1}$. 
Within 8 Mpc of us, 11 of 19 galaxies with $V_{\rm circ} > 150$ km~s$^{-1}$ show
no evidence for a classical bulge, one may contain a classical plus a 
pseudo bulge, and 7 are ellipticals or have classical bulges.  So it is hard to 
understand how bulgeless galaxies could form as the quiescent tail of a 
distribution of merger histories. \par
      Our second theme is environmental secular evolution.  We confirm that spheroidal 
galaxies have fundamental plane correlations that are almost perpendicular to those for 
bulges and elliptical galaxies.  Spheroidals are not dwarf ellipticals.  Rather, their 
structural parameters are similar to those of late-type galaxies.  We suggest that 
spheroidals are defunct late-type galaxies transformed by internal processes such as 
supernova-driven gas ejection and environmental processes such as secular harassment 
and ram-pressure stripping. \par
      Minus spheroidals, the fundamental plane correlations for ellipticals and bulges
have small scatter.  With respect to these, pseudobulges are larger and less dense.  
They fade out by becoming fluffy, not by becoming compact, like nuclei.  Pseudobulges 
and nuclear star clusters appear to have different origins.
\end{abstract}


\setcounter{page}{297}

\section{Internal and Environmental Secular Evolution}  

     Internal, slow (secular) evolution of galaxy disks occurs when 
nonaxisymmetries such as bars and spiral structure redistribute energy and angular 
momentum and rearrange disk structure.  Environmentally driven evolution can also
be secular (e.{\thinspace}g., galaxy harassment), although better known processes
are rapid \hbox{(mergers).} We concentrate on one consequence of environmental secular evolution.
It is one of several processes that can transform late-type dwarfs into ``spheroidals'',
i.{\thinspace}e., galaxies that are morphologically similar to ellipticals 
but that have different structural parameter correlations indicative of different 
formation physics.  \hbox{Figure 1} puts these galaxy formation processes into a more 
general context.

\begin{figure}[h!]
\pretolerance=15000  \tolerance=15000 
\plotfiddle{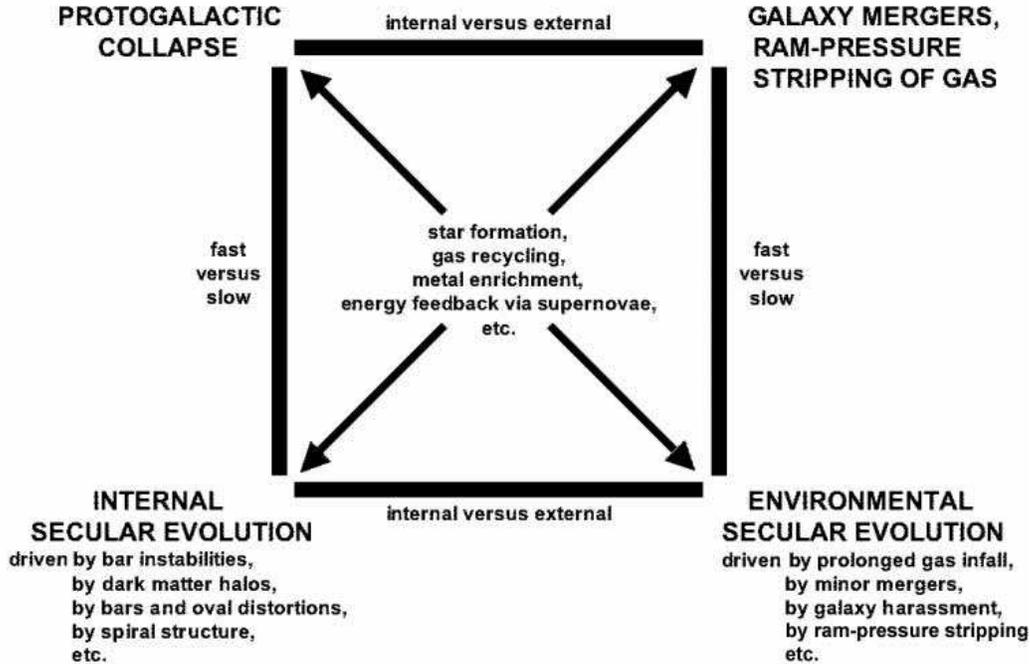}{3.2truein}{0}{68}{68}{-210}{-20}
\caption{\lineskip=-6pt \lineskiplimit=-6pt
Summary of galactic evolution processes (Kormendy \& Kennicutt 2004: KK04).
Processes are divided 
vertically into fast (top) and slow (bottom).  Fast evolution happens on a 
free-fall timescale, $t_{\rm ff} \sim (G\thinspace\rho)^{-1/2}$, where $\rho$ 
is volume density and $G$ is the gravitational constant.  Slow means 
many galaxy rotation periods.  Ram-pressure stripping is likely to be
fast for dwarf galaxies and slow for giant galaxies.  Processes are divided
horizontally into ones that happen internally in one galaxy (left) and ones 
that are driven by environmental effects such as galaxy interactions (right).
The processes at center affect all types of galaxy evolution.  This
paper reviews internal secular evolution in galaxy disks (lower-left) and
the nature of spheroidal galaxies as defunct late-type galaxies transformed
(right) by galaxy harassment, ram-pressure stripping, and other processes.}
\end{figure}

\section{Internal Secular Evolution and the Growth of Pseudobulges} 

\pretolerance=15000  \tolerance=15000 

      Aspects of internal secular evolution have long been thriving
``cottage industries'' (an early review is in Kormendy 1982).
Kormendy \& Kennicutt (2004) provide a synthesis of these many lines
of research, both observational and theoretical.  Other reviews are in
Sellwood \& Wilkinson (1993), Kormendy (1993), Buta \& Combes (1996)
Kormendy \& Cornell (2004), Kormendy \& Fisher (2005), Athanassoula (2007),
Peletier (2008), and Combes (2007, 2008).  With limited space, 
this paper concentrates on new observations of pseudobulge properties.  

      Whatever the engine, internal evolution has similar consequences.
Like all self-gravitating systems, galaxy disks tend to spread -- the outsides 
expand and the insides contract (Tremaine 1989).  This is as fundamental to disk 
evolution as core collapse is to globular clusters, as the production of hot 
Jupiters and colder Neptunes is to the evolution of planetary systems, and 
as evolution to red giants containing proto-white-dwarfs is to stellar evolution
(Kormendy \& Fisher 2005; Kormendy 2008).  In galaxy disks, gas infall and 
star formation builds dense central components that get mistaken for 
bulges but that were not made by galaxy mergers.  They come in several varieties 
depending on what drives the evolution.  Pseudobulges made from disk gas are often
but not always disky (Kormendy 1993; KK04; Fisher
\& Drory 2008a).  Box-shaped bulges also are disk phenomena:~they
are parts of edge-on bars 
(Combes \& Sanders 1981;
Combes et al.~1990;
Pfenniger \& Friedli 1991;
Raha et al.~1991;
Kuijken \& Merrifield 1995;
Merrifield \& Kuijken 1999;
Bureau \& Freeman 1999;
Bureau, Freeman, \& Athanassoula 1999;
Athanassoula 2005, 2007).  Nuclear bars are connected with disky pseudobulges 
(they rotate rapidly) and may be a subset of them.  Other morphology that 
identifies pseudobulges includes nuclear star formation rings and spiral structure.  
It is convenient to have one name -- ``pseudobulges'' --  for all central, 
high-density products of disk secular evolution.  

      How to identify pseudobulges is discussed in KK04.  Prototypical examples that 
are more disky than classical bulges were first recognized by their rapid rotation 
(Figure 2).  Disks have large $V_{\rm max}/\sigma$ and plot above the oblate line
when seen at inclinations other than edge-on. Early identification of very disky
(e.{\thinspace}g., NGC~4736, NGC 3945) and moderately disky (e.{\thinspace}g., NGC 2950, 
also a nuclear bar) pseudobulges have recently been augmented as shown in Figure 2. 

\begin{figure}[hb!]
\plotfiddle{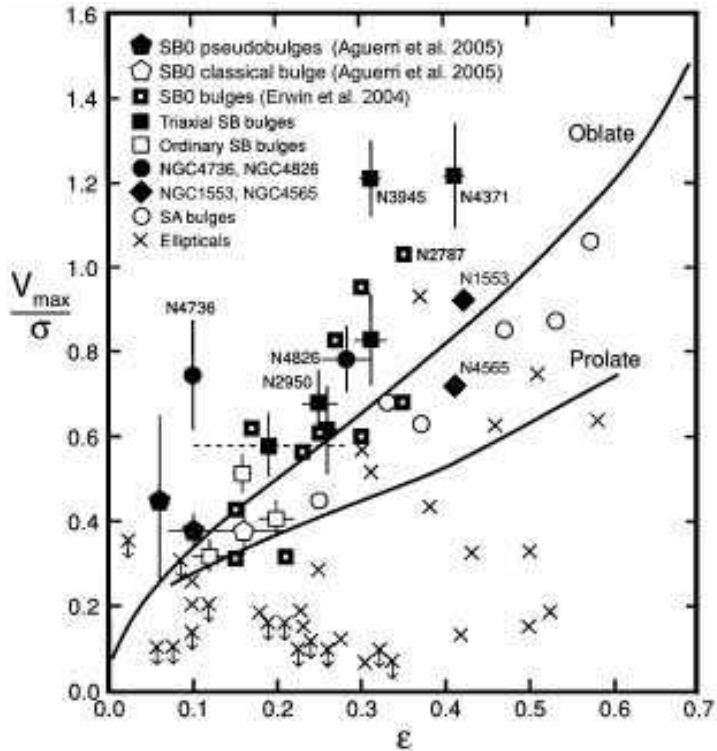}{3.5truein}{0}{47}{47}{-150}{-18}
\caption{Relative importance of rotation and random velocity as a function 
of observed ellipticity $\epsilon$ = (1\thinspace$-${\thinspace}axial ratio) for various kinds of
stellar systems.  Here $V_{\rm max}/\sigma$ is the ratio of maximum rotation
velocity to mean velocity dispersion interior to the half-light radius.  The 
``oblate'' line approximately describes oblate-spheroidal systems that have
isotropic velocity dispersions and that are flattened only by rotation; 
it is a consequence of the tensor virial theorem (Binney \& Tremaine 1987).  
This figure is updated from KK04.}
\end{figure}
\eject

     Figure 2 is an approximate analysis based on long-slit, major-axis spectra.  
Integral-field spectroscopy from the SAURON team now provides beautifully detailed 
detections of rapidly rotating, disky pseudobulges.  Often (Figure 3), it is 
exactly the high-surface-brightness center -- where the projected 
brightness profile rises above the inward extrapolation of the outer disk profile -- 
that shows rapid rotation and a corresponding inward decrease in velocity dispersion.  
Many of these kinematically decoupled components are also younger than the rest of 
the inner galaxy.  Some counter-rotate (McDermid et al.~2006) and presumably are 
made from accreted material.  But the phenomenon is common in barred and oval galaxies 
in which secular evolution is expected to be rapid.  Besides NGC 4274 in Figure 3, 
excellent examples include NGC 3623 (SABa in the optical but clearly SB(r) in 2MASS
$J${\null}$H${\null}$K$ images:~Jarrett et al.~2003), and NGC 5689 (SB0).  These
results are discussed in 
Ganda et al.~(2006),
Falc\'on-Barroso et al.~(2006),
Peletier et al.~(2007a);
see Peletier (2008) and
Peletier et al.~(2007b,{\thinspace}c) for reviews.  Quoting Peletier 
et al.~(2007c):~``SAURON observations show that 13 out of 24 Sa and Sab galaxies 
[and a similar fraction of late-type spirals] 
show a central local minimum in the velocity dispersion \dots~The sigma-drops are 
probably due to central disks that formed from gas falling into the central regions 
through a secular evolution process.''

\begin{figure}[hb!]
\def\gapprox{$_>\atop{^\sim}$} 
\def\lapprox{$_<\atop{^\sim}$}
\plotfiddle{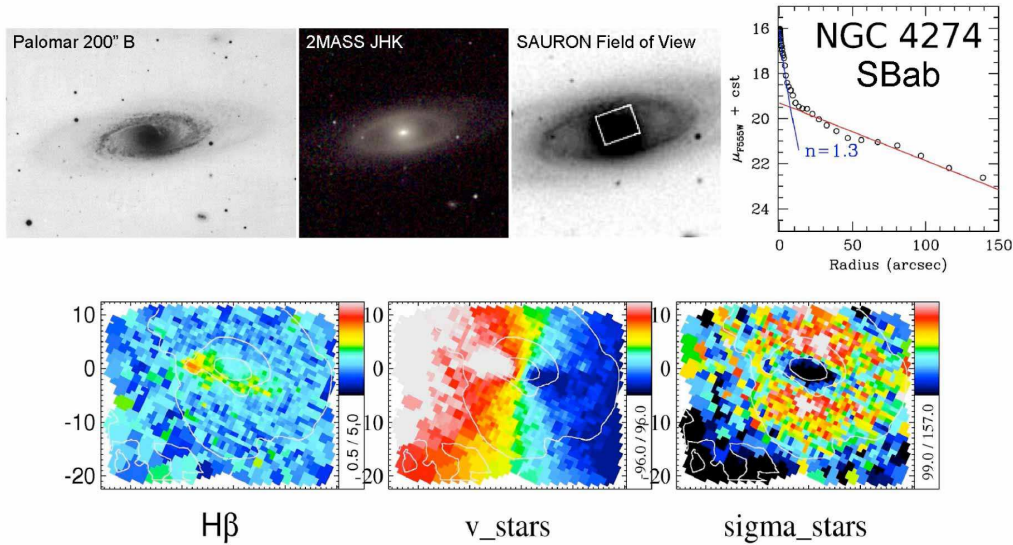}{2.5truein}{0}{67}{67}{-203}{-20}
\caption{\lineskip=0pt \lineskiplimit=0pt
SAURON integral-field spectroscopy of the disky pseudobulge in the Sa
galaxy NGC 4274 (adapted from Peletier et al.~2007c).  The images show that NGC 4274 
is a highly inclined barred galaxy; the bar is foreshortened, because it is
oriented nearly along the minor axis.  It fills an inner ring, as is normal in
SB(r) galaxies (Kormendy 1979).  The brightness profile (upper-right)
is decomposed into a S\'ersic (1968) function plus an exponential disk.  The
S\'ersic function has $n = 1.3$, i.{\thinspace}e., $n < 2$, as in other pseudobulges
(Figure 5).  The pseudobulge dominates the light at radii $r$~\lapprox 
\thinspace10$^{\prime\prime}$.  The kinematic maps (Falc\'on-Barroso et al.~2006)
show that this light comes from a disky component that is more rapidly rotating (center), 
lower in velocity dispersion (right), and stronger in H$\beta$ line strength (left,
from Peletier et al.~2007a) and hence younger than the rest of the inner galaxy.
}
\end{figure}
\eject

\begin{figure}[ht!]
\def\gapprox{$_>\atop{^\sim}$} 
\def\lapprox{$_<\atop{^\sim}$}
\plotfiddle{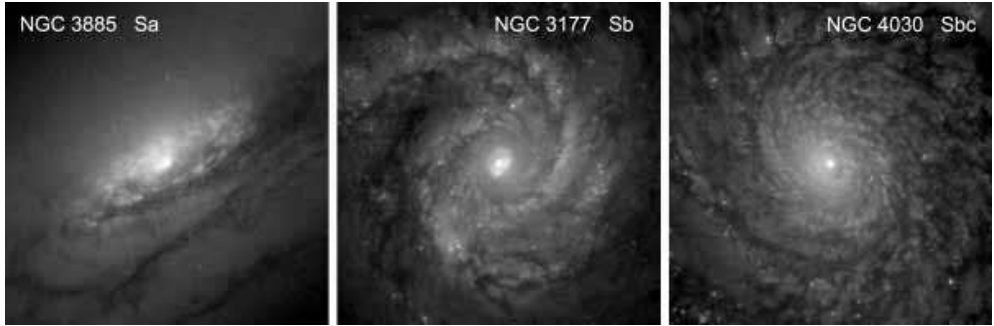}{1.5truein}{0}{65}{65}{-198}{-15}
\caption{\lineskip=0pt \lineskiplimit=0pt
Sa -- Sbc galaxies with disky pseudobulges shown in 18$^{\prime\prime}$ 
$\times$ 18$^{\prime\prime}$ regions centered on the galaxy nucleus and extracted from
{\it Hubble Space Telescope\/} (HST) WFPC2 F606W images kindly provided by 
Carollo et al. (1997, 1998).  Displayed intensity is proportional to the logarithm of the 
surface brightness.  Mean and minimum (pseudo)bulge-to-total luminosity 
ratios $B/T$ observed by Simien \& de Vaucouleurs (1986) are 0.35 and 0.13 for Sas, 0.22 and
0.10 for Sbs, and 0.18 and 0.05 for Sbcs.  These galaxies contain nuclear star clusters 
but no E-like component with the above $B/T$ values.
}
\end{figure}

      Pseudobulges were also recognized photometrically in Kormendy (1993); progress since
then has been rapid (KK04).  In many galaxies, the ``bulge'' is essentially as flat
as the disk and/or shows clearcut spiral structure.  Both are signatures of high-density
disks -- classical bulges are dynamically hot and cannot have small-scale spiral structure.  
These features are spectacular in HST surveys of the centers of spiral galaxies (Carollo 
et al.~1997, 1998, 2001, 2002; Carollo 1999).  
Figure 4 shows examples.  These are Sa{\thinspace}--{\thinspace}Sbc galaxies, so they should 
contain substantial bulges.  Instead, their centers look like star-forming spiral galaxies.  
Contrast the definition of a classical bulge (Renzini 1999 following Sandage 1961): A bulge is 
nothing more nor less than an elliptical galaxy that happens to live in the middle of a disk.

      Classical and pseudo bulges can coexist (KK04; Kormendy et al.\ 2006; Erwin 2007),
but the morphology in Figure 4 is not due to nuclear disks embedded in classical 
bulges that are hidden by the display parameters.  Bulges have steep brightness profiles, 
so bulge light would dilute the contrast in the spiral structure very strongly at smaller 
radii. But the strength of the spiral structure depends little on radius:~essentially all of the
pseudobulge participates.

     The imaging survey authors generally interpret disky bulges as consequences of secular
evolution.  Courteau, de Jong, \& Broeils (1996) observe ``spiral structure continuing into 
the central regions'' and ``invoke secular dynamical evolution  and \dots{\thinspace}gas 
inflow via angular momentum transfer and viscous transport'' as the explanation.  Carollo
et al.~(2001) conclude that ``exponential-type bulge formation is taking place in the local 
universe and that this process is consistent with being the outcome of secular evolution
\dots{\thinspace}within the disks''.

      A ``proof of concept'' is largely in hand, so studies of secular evolution now concentrate
on star formation (reviewed in Fisher \& Drory 2008b) and on pseudobulge statistical properties.  
In \S\thinspace4, we compare the fundamental plane parameter correlations of pseudobulges, 
classical bulges, and ellipticals.  First, we need to define what we mean by an elliptical.
This leads to our second theme on environmental secular evolution and the formation of 
spheroidal galaxies (\S\thinspace3).

\section{Environmental Secular Evolution: Origin of Spheroidal Galaxies} 

\pretolerance=15000  \tolerance=15000 

      Figure 5 shows fundamental plane projections 
(Djorgovski \& Davis 1987; 
Faber et al.~1987;
Djorgovski et al.~1988) and S\'ersic index versus total magnitude for various kinds of
stellar systems (adapted from Kormendy et al.~2008: KFCB).

\begin{figure}[hb!]
\def\gapprox{$_>\atop{^\sim}$} 
\def\lapprox{$_<\atop{^\sim}$}
\plotfiddle{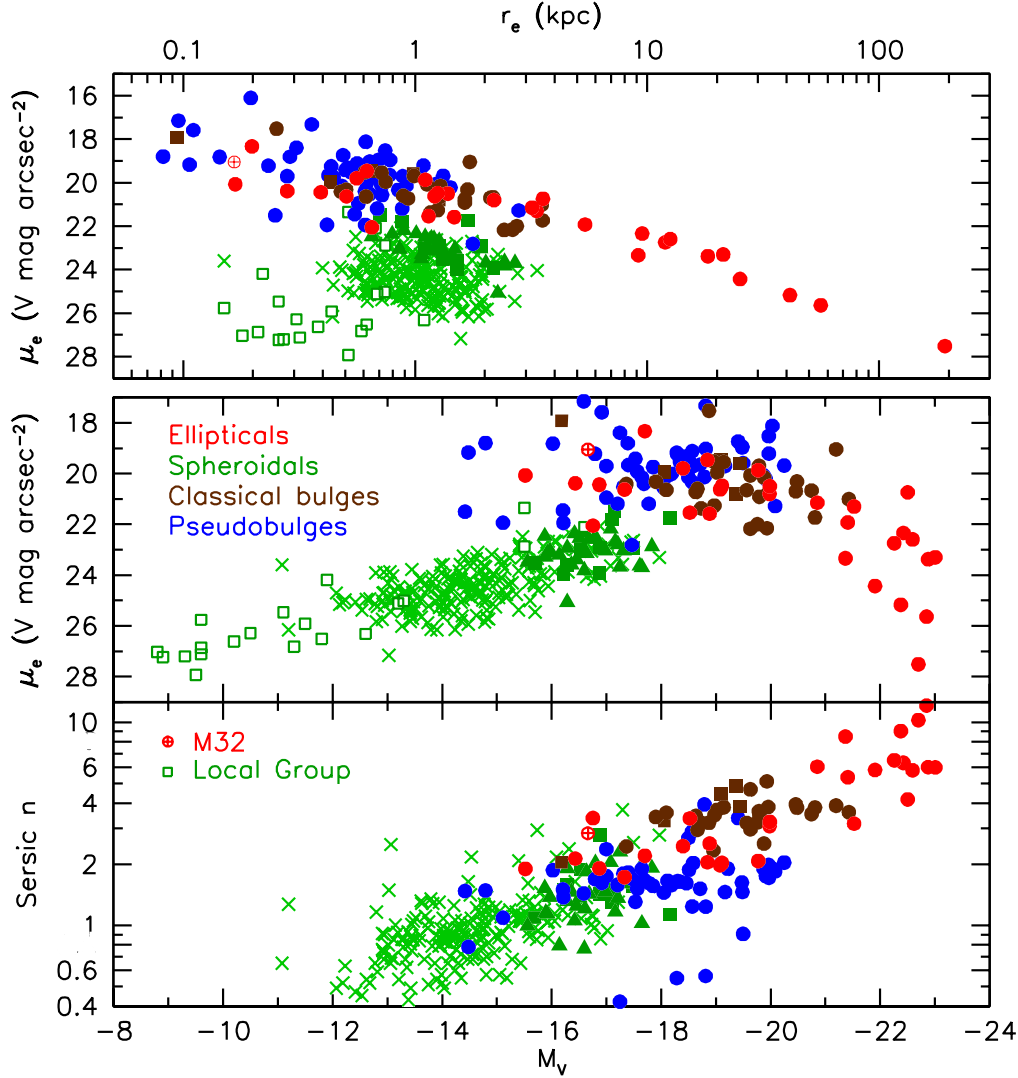}{5.2truein}{0}{82}{82}{-294}{-172}
\caption{\lineskip=0pt \lineskiplimit=0pt
Global parameter correlations for pseudobulges (blue), classical bulges (brown),
ellipticals (red), and spheroidal galaxies (green). Pseudobulge and most bulge 
points are from Fisher \& Drory (2007a).  The ellipticals, five bulge points, and 
the green squares are from Kormendy et al.~(2008:~KFCB).  Green triangles show 
all spheroidals from Ferrarese et al.~(2006) that are not in KFCB.  Crosses show 
all spheroidals from Gavazzi et al.~(2005) that are not in KFCB or in Ferrarese 
et al.~(2006).  Open squares are Local Group spheroidals (Mateo 1998; McConnachie 
\& Irwin 2006).  The bottom panels show major-axis S\'ersic index $n$ and 
effective surface brightness $\mu_e$ versus total galaxy absolute magnitude.  
The top panel shows $\mu_e$ vs.~effective radius $r_e$ (the Kormendy 1977 
relation, which shows the fundamental plane almost edge-on).
}
\end{figure}

\eject

\def\gapprox{$_>\atop{^\sim}$} 
\def\lapprox{$_<\atop{^\sim}$}

\noindent Critical to the interpretation of this figure is high-accuracy photometry of
all known E and selected Sph galaxies in the Virgo cluster from KFCB.  Composite HST and 
ground-based profiles over large radius ranges provide accurate S\'ersic parameters.
Then the intrinsically small scatter of the fundamental plane (Saglia et al.~1993; 
J\o rgensen et al.~1996) is seen in the top panel, which shows the plane almost edge-on.  
Figure 5 confirms the results of Kormendy (1985, 1987), Binggeli \& Cameron (1991), and 
Bender, Burstein, \& Faber (1992) that E and Sph galaxies satisfy different parameter 
correlations.  This result has been criticized by Jerjen \& Binggeli (1997), 
Graham \& Guzm\'an (2003), Gavazzi et al.~(2005),
and Ferrarese et al.~(2006) in part because the $n$ -- $M_V$ correlation is continuous.  
We agree.  But the observation that $n$ is not sensitive to the difference between E and Sph
galaxies does not mean that they are related.  The fundamental plane correlations (top panels
and Figure 6) show that lower-luminosity Es are monotonically higher in density, whereas
lower-luminosity Sphs are monotonically lower in density.  Spheroidals are not
faint ellipticals.  Instead, Kormendy (1985, 1987) showed that they have similar parameter
correlations to dwarf spiral and irregular galaxies.  Spheroidals and ellipticals almost
certainly had very different formation processes. We believe that Es formed via major
galaxy mergers.  Evidence discussed in KFCB suggests that spheroidal galaxies are defunct 
late-type galaxies transformed by internal processes such as supernova-driven gas ejection
(Dekel \& Silk 1986) and environmental processes such as secular galaxy harassment
(Moore et al.~1996, 1998) and ram-pressure gas stripping (e.{\thinspace}g., Chung et al.~2008).

\vfill

\begin{figure}[hb!]
\def\gapprox{$_>\atop{^\sim}$} 
\def\lapprox{$_<\atop{^\sim}$}
\plotfiddle{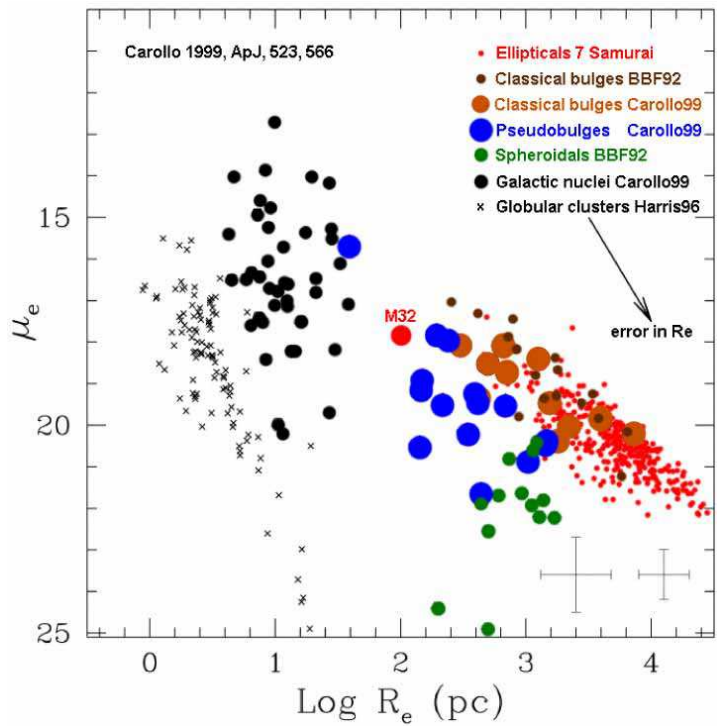}{1truein}{0}{47}{47}{-150}{-20}
\caption{\lineskip=0pt \lineskiplimit=0pt
Effective surface brightness versus effective radius for various kinds of stellar 
systems (adapted from Carollo 1999; colors as in Figure 5).   
}
\end{figure}
\eject

\section{Fundamental Plane Correlations for Bulges and Pseudobulges} 

\pretolerance=15000  \tolerance=15000 

      Figures 5 and 6 compare parameter correlations for ellipticals, classical bulges,
and pseudobulges.  S\'ersic parameters of (pseudo)bulges are less accurate than those 
of Es, because deriving them requires a decomposition of brightness profiles into 
(pseudo)bulge and disk contributions.  We have reduced leverage on bulge parameters, 
and they are strongly coupled to the disk parameters.  
Nevertheless, Figures 5 and 6 show that classical bulges are indeed indistinguishable 
from elliptical galaxies, consistent with our definition.  Many pseudobulges are not very 
different, either; this is one reason why they got confused with bulges.  To find the
difference between pseudobulges and classical bulges, we need to look beyond
fundamental plane parameters and consider properties such as flattening and $V/\sigma$.
Nevertheless, in Figures~5~and~6, pseudobulges also show larger scatter than classical 
bulges, and they have smaller S\'ersic indices.  Consistent with Courteau et al.~(1996),
MacArthur, Courteau, \& Holtzman (2003), and the Carollo team papers, Fisher \& Drory (2007a) 
find a relatively clean separation between classical bulges with $n$ \gapprox \thinspace\thinspace2
and pseudobulges (mostly) with $n$ \lapprox \thinspace\thinspace2.  Note that this conclusion
would not be clear if we believed that spheroidals are faint ellipticals.  In the above,
the bulge-pseudobulge distinction is based on morphological criteria listed in KK04 and 
not on profile shape.  We do not understand galaxy formation well enough to predict $n$
for either type of bulge, but the distinction is clearcut enough to be a classification 
aid. 

      Figure 6 shows that pseudobulges fade out by becoming low in density, not by
becoming compact, like nuclear star clusters (black filled circles).  This suggests
that pseudobulges and nuclei are fundamentally different.

\section{How can hierarchical clustering make so many bulgeless galaxies?} 

      Hierarchical clustering in a cold dark matter universe (White \& Rees 1978) is
a remarkably successful theory of galaxy formation.  The struggle now is with baryonic
physics.  The most serious problem has been emphasized many times, both by observers 
(e.{\thinspace}g., Freeman 2000; KK04; Kormendy \& Fisher 2005; Carollo et al.~2007; 
Kormendy 2008) and by modelers (e.{\thinspace}g., Steinmetz \& Navarro 2002; 
Abadi et al.~2003).  Given so much merger violence, 
how can hierarchical clustering produce so many pure disk galaxies with no signs of 
merger-built bulges?  This problem gets harder when we realize that many of what we used
to think are small bulges are really pseudobulges made by secular evolution.  We know of
no Sc or later-type galaxy with a classical bulge (KK04).  So the solution to the
above problem is not to hope that bulgeless disks are rare enough that they can be
explained as the tail of a distribution of formation histories that included a few
fortuitously mergerless galaxies.

      This section provides new examples and better statistics on bulgeless disks.

      The bulgeless disks that most constrain our formation picture are those that live
in high-mass dark halos -- say, ones in which circular-orbit rotation velocities are 
$V_{\rm circ} \sim 200$ km s$^{-1}$.  Kormendy et al.~(2009) have used the Hobby-Eberly 
Telescope to obtain high-resolution (instrumental dispersion $\sigma_{\rm instr} \simeq 8$ 
km s$^{-1}$) spectroscopy of the nuclear star clusters in M{\thinspace}101 and NGC 6946.  
M{\thinspace}101 is an Scd galaxy with $V_{\rm circ} = 210 \pm 15$ km s$^{-1}$
(Bosma et al.~1981).  But its nucleus has a velocity dispersion 
$\sigma = 25 \pm 7$ km s$^{-1}$ like that of a big globular cluster.  NGC 6946 is a similar 
Scd with $V_{\rm circ} = 210 \pm 10$ km s$^{-1}$ (Tacconi \& Young 1986; Sofue 1996) 
and $\sigma = 38 \pm 3$ km s$^{-1}$.  IC 342 is a third such galaxy with 
$V_{\rm circ} = 192 \pm 5$ km s$^{-1}$ (Rogstad, Shostak, \& Rots 1973; Sofue 1996) and 
$\sigma = 33 \pm 3$ km s$^{-1}$ (B\"oker et al.~1999; 
$\sigma_{\rm instr} = 5.5$ km s$^{-1}$).  All three galaxies show small central upturns in
their $J$\null$H$\null$K$ brightness profiles (Jarrett et al.~2003) and NGC 6946 and IC 342
also show rapid rises in their central CO rotation curves $V(r)$ (Sofue 1996).  But their 
small dispersions $\sigma \ll V$  show that these are pseudobulges.  How did these halos 
grow so large with no signs of major mergers?

      Could bulgeless disks be rare enough to have formed as the quiescent tail of a 
distribution of merger histories?  We believe that the answer is ``no''.  Consider first 
the Local Group.  Only our Galaxy has uncertainty in its bulge classification.  The 
box-shaped structure implies a pseudobulge.  The low velocity dispersion of the bulge 
merges seamlessly with that of the disk (Lewis \& Freeman 1989).  The central $\sigma$
profile derived by Tremaine et al.~(2002) implies a pseudobulge.  
Only the old, $\alpha$-element-enhanced stellar population is suggestive of a classical 
bulge (KK04 discusses these caveats).  In agreement with Freeman (2008), we conclude that
there is no photometric or dynamical evidence for a classical bulge.  Then 
the Local Group contains one elliptical, M{\thinspace}32, and one classical bulge, 
in M{\thinspace}31.  In the most massive three galaxies, there is only one classical bulge.

      Looking beyond the Local Group, the most distant bulgeless disk discussed above is 
M{\thinspace}101.  Its Cepheid distance modulus is $m - M = 29.34 \pm 0.10$; i.{\thinspace}e., 
distance = $7.4 \pm 0.3$ Mpc (Ferrarese et al.~2000), and $V_{\rm circ} = 210 \pm 15$ 
km s$^{-1}$.  We will be conservative and look for all galaxies with $V_{\rm circ} > 150$ 
km s$^{-1}$ or central $\sigma > 106$ km s$^{-1}$ and $m - M < 29.5$.  HyperLeda and
Tonry et al.~(2001) provide 19 such galaxies.  M{\thinspace}101, NGC 6946, and IC 342, 
are 3/19 of the big galaxies in our sample volume.  Of the rest, 8 are dominated by
pseudobulges with no sign of a classical bulge.  One more, NGC 2787, has a dominant 
pseudobulge but could also have a small classical bulge component.  Three galaxies in 
the above volume are ellipticals, Maffei 1, NGC 3077 (probably), and NGC 5128.  Three 
galaxies are known to have classical bulges, M{\thinspace}31, M{\thinspace}81, and NGC 4258.  
NGC 5195, the companion of M{\thinspace}51, has an uncertain classification but 
$\sigma = 157$ km s$^{-1}$; we include it among the classical bulges.  This leaves us 
with the following statistics: Within 8 Mpc of us, 11 of 19 galaxies with 
$V_{\rm circ} > 150$ km  s$^{-1}$ show no evidence for a classical bulge, one may contain 
both a classical bulge and a pseudobulge, and 7 of 19 are either ellipticals or contain 
classical bulges. Big galaxies with evidence for a major merger are less than half of the sample.

      In contrast, in the Virgo cluster, about 2/3 of the stellar mass is in
elliptical galaxies and some additional mass is in classical bulges (KFCB).  So
the above statistics are a strong function of environment.

      We therefore restate the theme of this section: What is special about 
galaxy formation in low-density, Local-Group-like environments that allows $>$\thinspace1/2
of the galaxies with halo $V_{\rm circ} > 150$ km s$^{-1}$ to form
with no signs of major mergers?

\acknowledgements We thank Ralf Bender, Mark Cornell, Niv Drory and Reynier Peletier
for permission to quote results before publication.  Our work used the HyperLeda database at 
{\tt http://leda.univ-lyon1.fr/search.html}.  Support from the National Science Foundation 
under grant AST-0607490 is gratefully acknowledged.


\begin{thebibliography}{}

\bibitem[]{} Abadi, M.~G., Navarro, J.~F., Steinmetz, M., \& Eke, V.~R.~2003, ApJ, 591, 499

\bibitem[]{} Aguerri,{\thinspace}J.{\thinspace}A.{\thinspace}L., Elias-Rosa,{\thinspace}N., 
             Corsini,{\thinspace}E.{\thinspace}M., \& Mu\~noz-Tu\~n\'on,{\thinspace}C.{\thinspace}2005, A\&A, 434, 109

\bibitem[]{} Athanassoula, E.~2005, MNRAS, 358, 1477

\bibitem[]{} Athanassoula, E.~2007, in Mapping the Galaxy and Nearby Galaxies, ed.~K.~Wada \& 
             F.~Combes (New York: Springer), in press (astro-ph/0610113) 

\bibitem[]{} Bender, R., Burstein, D., \& Faber, S.~M.~1992, ApJ, 399, 462

\bibitem[]{} Binggeli, B., \& Cameron, L.~M.~1991, A\&A, 252, 27

\bibitem[]{} Binney, J., \& Tremaine, S.~1987, Galactic Dynamics (Princeton:
    Princeton Univ.~Press)

\bibitem[]{} B\"oker, T., van der Marel, R.~P., \& Vacca, W.~D.~1999, AJ, 118, 831

\bibitem[]{} Bosma, A., Goss, W.~M., \& Allen, R.~J.~1981, A\&A, 93, 106

\bibitem[]{} Bureau, M., \& Freeman, K.~{\thinspace}C.~1999, AJ, 118, 126

\bibitem[]{} Bureau, M., Freeman, K.~{\thinspace}C., \& Athanassoula, E.~1999, in The
             Formation of Galactic Bulges, ed.~C.{\thinspace}M.~Carollo et al.~(Cambridge: 
             Cambridge Univ. Press), 115

\bibitem[]{} Buta, R., \& Combes, F.~1996, Fund.~Cosm.~Phys.,~17, 95

\bibitem[]{} Carollo, C.~M.~1999, ApJ, 523, 566

\bibitem[]{} Carollo, C.~M., Stiavelli, M., de Zeeuw, P.~T., \& Mack, J.~1997, AJ, 114, 2366

\bibitem[]{} Carollo, C.~M., Stiavelli, M., \& Mack, J.~1998, AJ, 116, 68

\bibitem[]{} Carollo,~C.~M., et al.~2002, AJ, 123, 159

\bibitem[]{} Carollo, C.~M., et al.~2001, ApJ, 546, 216

\bibitem[]{} Carollo, C.~M., et al.~2007, ApJ, 658, 960

\bibitem[]{} Chung, A.~et al.~2008, in Formation and Evolution of Galaxy Disks, ed.~J.~G.~Funes, 
             S.{\thinspace}J.~\& E.~M.~Corsini (San Francisco: ASP), 127

\bibitem[]{} Combes, F.~2007, in IAU Symposium 245, Formation and Evolution of Galaxy
             Bulges, ed.~M.~Bureau et al.~(Cambridge: Cambridge University Press), in press

\bibitem[]{} Combes, F.~2008, in Formation and Evolution of Galaxy Disks, ed.~J.~G.~Funes, 
             S.{\thinspace}J.~\& E.~M.~Corsini (San Francisco: ASP), 325

\bibitem[]{} Combes, F., Debbasch, F., Friedli, D., \& Pfenniger, D.~1990, A\&A, 233, 82

\bibitem[]{} Combes, F., \& Sanders, R.~H~1981, A\&A, 96, 164

\bibitem[]{} Courteau, S., de Jong, R.~S., \& Broeils, A.~H.~1996, ApJ, 457, L73

\bibitem[]{}  Dekel, A., \& Silk, J.~1986, ApJ, 303, 39

\bibitem[]{} Djorgovski, S., \& Davis, M.~1987, ApJ, 313, 59

\bibitem[]{} Djorgovski, S., de Carvalho, R., \& Han, M.-S.~1988, in The Extragalactic 
             Distance Scale, ed.~S.~van den Bergh \& C.~J.~Pritchet (San Francisco: ASP), 329

\bibitem[]{} Erwin, P.~2007, in IAU Symposium 245, Formation and Evolution of Galaxy
             Bulges, ed.~M.~Bureau et al.~(Cambridge: Cambridge Univ.~Press), in press

\bibitem[]{} Erwin, P., Beckman, J.~E., \& Vega Beltran, J.~C.~2004, in Penetrating Bars
             Through Masks of Cosmic Dust, ed.~D.~L.~Block et al. (New York: Springer), 775

\bibitem[]{} Faber, S.~M., et al.~1987, in Nearly Normal Galaxies: From the Planck
             Time to the Present, ed. S.~M.~Faber (New York: Springer), 175 

\bibitem[]{} Falc\'on-Barroso, J., et al.~2006, MNRAS, 369, 529 

\bibitem[]{} Ferrarese, L., et al.~2000, ApJS, 128, 431

\bibitem[]{} Ferrarese, L., et al.~2006, ApJS, 164, 334 

\bibitem[]{} Fisher, D.~B., \& Drory, N.~2008a, AJ, 136, 773

\bibitem[]{} Fisher, D.~B., \& Drory, N.~2008b, in Formation and Evolution of Galaxy Disks, 
             ed.~J.~G. Funes, S.{\thinspace}J.~\& E.~M.~Corsini (San Francisco: ASP), 309

\bibitem[]{} Freeman, K.~C.~2000, in Toward a New Millennium in Galaxy Morphology,
     ed.~D.~L.~Block et al. (Dordrecht: Kluwer), 119

\bibitem[]{} Freeman, K.~C.~2008,  in Formation and Evolution of Galaxy Disks, ed.~J.~G.~Funes, 
             S.{\thinspace}J.~\& E.~M.~Corsini (San Francisco: ASP), 3

\bibitem[]{} Ganda, K., et al.~2006, MNRAS, 367, 46  

\bibitem[]{} Gavazzi, G., et al.~2005, A\&A, 430, 411

\bibitem[]{} Graham, A.~W., \& Guzm\'an, R.~2003, AJ, 125, 2936

\bibitem[]{} Jarrett, T.~H., et al.~2003, AJ, 125, 525

\bibitem[]{} Jerjen, H., \& Binggeli, B.~1997, in The Second Stromlo Symposium:
             The Nature of Elliptical Galaxies, ed.~M. Arnaboldi, et al.~(San Francisco:
             ASP), 239

\bibitem[]{} J\o rgensen, I., Franx, M., \& Kj\ae rgaard, P.~1996, MNRAS, 280, 167

\bibitem[]{} Kormendy, J.~1977, ApJ, 218, 333

\bibitem[]{} Kormendy, J.~1979, ApJ, 227, 714

\bibitem[]{} Kormendy, J., 1982, in Morphology and Dynamics of Galaxies, Twelfth
           Saas-Fee Course, 
           ed.~L.~Martinet \& M.~Mayor (Sauverny: Geneva Observatory), 113

\bibitem[]{} Kormendy, J.~1985, ApJ, 295, 73

\bibitem[]{} Kormendy, J.~1987, in Nearly Normal Galaxies: From the Planck Time
             to the Present, ed.~S.~M.~Faber (New York:~Springer), 163

\bibitem[]{} Kormendy, J.~1993, in IAU Symposium 153, Galactic Bulges, 
     ed.~H.~Dejonghe \& H.~J.~Habing (Dordrecht: Kluwer), 209

\bibitem[]{} Kormendy, J.~2008, in IAU Symposium 245, Formation and Evolution of Galaxy
     Bulges, ed.~M.~Bureau et al.~(Cambridge: Cambridge Univ.~Press), 107

\bibitem[]{} Kormendy, J., \& Cornell, M. E.~2004, in Penetrating Bars
      Through Masks of Cosmic Dust, ed.~D.~L.~Block et al. (New York: Springer), 261

\bibitem[]{} Kormendy, J., et al.~2006, ApJ, 642, 765

\bibitem[]{} Kormendy, J., Drory, N., Bender, R., \& Cornell, M.~E.~2009, ApJ, in preparation

\bibitem[]{} Kormendy, J., \& Fisher, D.~B.~2005, RevMexA\&A (Serie de Conferencias), 23, 101

\bibitem[]{} Kormendy, J., Fisher, D.~B., Cornell, M.~E., \& Bender, R.~2008, ApJS (arXiv:0810.1681)

\bibitem[]{} Kormendy, J., \& Kennicutt, R.~C.~2004, ARA\&A, 42, 603 (KK04)

\bibitem[]{} Kuijken, K., \& Merrifield, M.~R.~1995, ApJ, 443, L13

\bibitem[]{} Lewis, J.~R., \& Freeman, K.~C.~1989, AJ, 97, 139

\bibitem[]{} MacArthur L.~A., Courteau, S., \& Holtzman, J.~A.~2003, ApJ, 582, 689

\bibitem[]{} Mateo, M.~1998, ARA\&A, 36, 435

\bibitem[]{} McConnachie, A.~W., \& Irwin, M.~J.~2006, MNRAS, 365, 1263

\bibitem[]{} McDermid, R.~M.~et al.~2006, MNRAS, 373, 906 

\bibitem[]{} Merrifield, M.~R., \& Kuijken, K.~1999, A\&A, 345, L47 

\bibitem[]{} Moore, B., et al.~1996, Nature, 379, 613  

\bibitem[]{} Moore, B., Lake, G., \& Katz, N.~1998, ApJ, 495, 139  

\bibitem[]{} Peletier, R.~F.~2008, in Pathways Through an Eclectic Universe, ed.~J.~H.~Knapen,
             T.~J.~Mahoney, \& A.~Vazdekis (San Francisco: ASP), 232

\bibitem[]{} Peletier, R.~F., et al.~2007a, MNRAS, 379, 445 

\bibitem[]{} Peletier, R.~F., et al.~2007b, in IAU Symposium 241, Stellar Populations as Building 
             Blocks of Galaxies, ed.~A.~Vazdekis \& 
             R.~Peletier (Cambridge Univ.~Press), 485 

\bibitem[]{} Peletier, R.~F., et al.~2007c, in IAU Symposium 245, Formation and Evolution of Galaxy
             Bulges, ed.~M.~Bureau et al.~(Cambridge: Cambridge Univ.~Press), in press 

\bibitem[]{} Pfenniger, D., \& Friedli, D.~1991. A\&A, 252, 75

\bibitem[]{} Raha, N., Sellwood, J.~A., James, R.~A., \& Kahn, F.~D.~1991, Nature, 
             352, 411

\bibitem{} Renzini, A.~1999, in The Formation of Galactic Bulges, ed.~C.~M.~Carollo 
           et al. (Cambridge: Cambridge Univ.~Press), 9

\bibitem[]{} Rogstad, D.~H., Shostak, G.~S., \& Rots, A.~H.~1973, A\&A, 22, 111

\bibitem[]{} Saglia, R.~P, Bender, R., \& Dressler, A.~1993, A\&A, 279, 75

\bibitem[]{} Sandage, A.~1961, The Hubble Atlas of Galaxies (Carnegie Institution of Washington)


\bibitem[]{} Sellwood, J.~A., \& Wilkinson, A.~1993, Rep.~Prog.~Phys., 56, 173

\bibitem[]{} S\'ersic, J.~L.~1968, Atlas de Galaxias Australes (Cordoba: 
           Obs.~Astr., Univ.~de Cordoba)

\bibitem[]{} Simien, F., \& de Vaucouleurs, G.~1986, ApJ, 302, 564

\bibitem[]{} Sofue, Y.~1996, ApJ, 458, 120

\bibitem[]{} Steinmetz, M., \& Navarro, J.~F.~2002, NewA, 7, 155

\bibitem[]{} Tacconi, L.~J., \& Young, J.~S.~1986, ApJ, 308, 600

\bibitem[]{} Tonry, J.~L., et al.~2001, ApJ, 546, 681

\bibitem[]{} Tremaine S.~1989, in Dynamics of Astrophysical Disks, 
             ed.~J.~A.~Sellwood (Cambridge: Cambridge Univ. Press), 231

\bibitem[]{} Tremaine, S., et al.~2002, ApJ, 574, 740

\bibitem[]{} White, S.~D.~M., \& Rees, M.~J.~1978, MNRAS, 183, 341

\end{thebibliography}
\end{document}